\documentclass[preprint]{iau}

\usepackage{graphicx}

\newcommand{\degree}{\ensuremath{^\circ}}
\newcommand{\simi}{\ensuremath{\sim}}
\newcommand{\nd}{\ensuremath{^{\textrm{\scriptsize nd}}} }

\title[ASTEP South] 
{ASTEP~South: a first photometric analysis}

\author[Nicolas Crouzet et al.]   
{N. Crouzet$^1$, T. Guillot$^2$, D. M\'ekarnia$^2$, J. Szul\'agyi$^{2,3}$, L. Abe$^2$, A. Agabi$^2$, Y. Fante\"\i-Caujolle$^2$, I. Gon\c calves$^2$, M. Barbieri$^2$, F.-X. Schmider$^2$, J.-P. Rivet$^2$, E. Bondoux$^4$, Z. Challita$^4$, C. Pouzenc$^4$, F. Fressin$^5$, F. Valbousquet$^6$, A. Blazit$^2$, S. Bonhomme$^2$, J.-B. Daban$^2$, C. Gouvret$^2$, D. Bayliss$^7$, G. Zhou$^7$, and the ASTEP team}

\affiliation{$^1$Space Telescope Science Institute, 3700 San Martin Drive, Baltimore, MD 21218, USA \\email: {\tt crouzet@stsci.edu} \\[\affilskip]
$^2$Laboratoire Lagrange, UMR 7293 UNS-CNRS-OCA, Boulevard de l'Observatoire, BP 4229, 06304 Nice Cedex 4, France \\[\affilskip]
$^3$Konkoly Observatory, Research Centre for Astronomy and Earth Sciences, Hungarian Academy of Sciences, Konkoly Thege Mikl\'os \'ut 15-17, H-1121 Budapest, Hungary \\[\affilskip]
$^4$Concordia Station, Dome C, Antarctica \\[\affilskip]
$^5$Harvard-Smithsonian Center for Astrophysics, 60 Garden Street, Cambridge, MA 02138, USA \\[\affilskip]
$^6$Optique et Vision, 6 bis avenue de l'Est\'erel, BP 69, 06162 Juan-Les-Pins, France \\[\affilskip]
$^7$Research School of Astronomy \& Astrophysics, The Australian National University, Cotter Road, Weston Creek, ACT 2611, Australia}

\pubyear{}
\volume{288}  
\pagerange{}
\jname{Astrophysics from Antarctica}
\editors{M.G. Burton, X. Cui \& N.F.H. Tothill, eds.}
\begin{document}

\maketitle

\begin{abstract}

The ASTEP project aims at detecting and characterizing transiting planets from Dome~C, Antarctica, and qualifying this site for photometry in the visible. The first phase of the project, ASTEP South, is a fixed 10~cm diameter instrument pointing continuously towards the celestial South pole. Observations were made almost continuously during 4 winters, from 2008 to 2011. The point-to-point RMS of 1-day photometric lightcurves can be explained by a combination of expected statistical noises, dominated by the photon noise up to magnitude 14. This RMS is large, from 2.5 mmag at $R=8$ to 6\% at $R=14$, because of the small size of ASTEP South and the short exposure time (30 s). Statistical noises should be considerably reduced using the large amount of collected data. A 9.9-day period eclipsing binary is detected, with a magnitude $R=9.85$. The 2-season lightcurve folded in phase and binned into 1000 points has a RMS of 1.09 mmag, for an expected photon noise of 0.29 mmag. The use of the 4 seasons of data with a better detrending algorithm should yield a sub-millimagnitude precision for this folded lightcurve. Radial velocity follow-up observations are conducted and reveal a F-M binary system. The detection of this 9.9-day period system with a small instrument such as ASTEP South and the precision of the folded lightcurve show the quality of Dome C for continuous photometric observations, and its potential for the detection of planets with orbital period longer than those usually detected from the ground.

\end{abstract}

\section{Introduction}

Dome~C offers exceptional conditions for astronomy thanks to a 3-month continuous night during the Antarctic winter and a very dry atmosphere. This site is located at $\rm{75\degree 06'S - 123\degree 21'E}$ at an altitude of 3233 meters on a summit of the high Antarctic plateau, 1100 km away from the coast. The winter site testing for astronomy revealed a very clear sky, an excellent seeing above a thin boundary layer, very low wind-speeds (\cite{Aristidi2003,Aristidi2005a,Lawrence2004a,Ashley2005a,Aristidi2009,Giordano2012,Fossat2010}), a very low scintillation (\cite{Kenyon2006a}) and a high duty cycle (\cite{Mosser2007a,Crouzet2010}). Time-series observations such as those implied by the detection of transiting exoplanets should benefit from these atmospherical conditions and the good phase coverage (\cite{Pont2005a}).
The ASTEP project (Antarctic Search for Transiting ExoPlanets) aims at determining the quality of Dome~C as a site for future photometric surveys and to detect transiting planets (\cite{Fressin2005a}). The main instrument is a 40~cm Newton telescope entirely designed and built to perform high precision photometry from Dome~C. The design is presented in \cite{Daban2010} and the performances are detailed in \cite{Rivet2012}. A 10~cm instrument pointing continuously towards the celestial South pole, ASTEP South, was first installed at Dome~C (\cite{Crouzet2010}). Both instruments use facilities provided by the French-Italian Concordia station at Dome C, and are installed on the AstroConcordia platform at the ground level. Other photometric instruments are also observing from Antarctica. Among them is CSTAR, an instrument very similar to ASTEP South and located at Dome A (\cite{Yuan2008,Zhou2010a}). Variable stars in the South pole field were detected and classified using CSTAR, and a first planetary candidate was reported (\cite{Zhou2010b}, Wang \textit{et al.} 2011). Here, we present a photometric analysis of the ASTEP South data. First, we briefly describe the instrumental setup. Then, we detail the lightcurve extraction and show its performance over 1 day. Finally, we present our independent detection of the planetary candidate reported by \cite{Wang2011}, as well as follow-up observations.

\section{Observations}

ASTEP South consists of a 10~cm refractor, a front-illuminated 4096x4096 pixels CCD camera, and a simple mount in a thermalized enclosure. The refractor is a commercial TeleVue NP101. The camera is a ProLines series by Finger Lake Instrumentation equipped with a KAF-16801E CCD by Kodak (see \cite{Crouzet2007} for the choice of the camera). The overall transmission (600 to 900\,nm) is equivalent to that of a large R band. The enclosure is thermalized to $-$20\degree C and closed by a double glass window on the optical path to avoid temperature fluctuations. The instrument is shown in Figure~\ref{fig:ASTEPSouthDomeC}. ASTEP South is completely fixed and points towards the celestial South pole continuously. The observed field of view is $3.88\times 3.88\degree$$^2$, leading to a pixel size of 3.41 arcsec on the sky. This field contains around 8000 stars up to magnitude $R$~=~15. This observation setup leads to stars moving circularly on the CCD with a 1-sidereal day period, and to an elongated PSF (Point Spread Function). The exposure time is 30 s with 10 s between each exposure. The PSF is defocused to a 2 px FWHM (Full Width Half Maximum), but varies in particular with the seeing at the ground-level where the instrument is placed. ASTEP South observed during 4 winters:  from June to the end of the winter in 2008 collecting \simi1500 hours of data, and during all winter in 2009, 2010, and 2011 collecting \simi2500 hours of data per winter. 


\section{Photometric analysis}

Due to the rotation of the stars on the CCD, the PSFs are elongated up to 4.5 px at the edges of the field, with an elongation direction varying in time. We designed a specific photometric algorithm to take into account this particular feature: an elongated photometric aperture is created for each star in each image. The aim is to reduce the number of pixels compared to standard circular apertures in order to minimize the sky background and read-out noise. The aperture elongation and orientation are calculated according to the position of the star on the image. The size is optimized empirically for each star. To this end, we perform photometry on a particular day (June 21, 2008) using a large number of aperture sizes. The size yielding the lower point-to-point RMS is kept. As expected, bright stars end up with a larger aperture than faint stars. The aperture size of each star then remains constant during the process. In particular, it does not vary with the seeing: we found that a greater average seeing does not necessarily yield a larger aperture when performing the optimization.



\begin{figure}[htbp]
\begin{minipage}[htbp]{5cm}
   \centering
   \includegraphics[width=5cm]{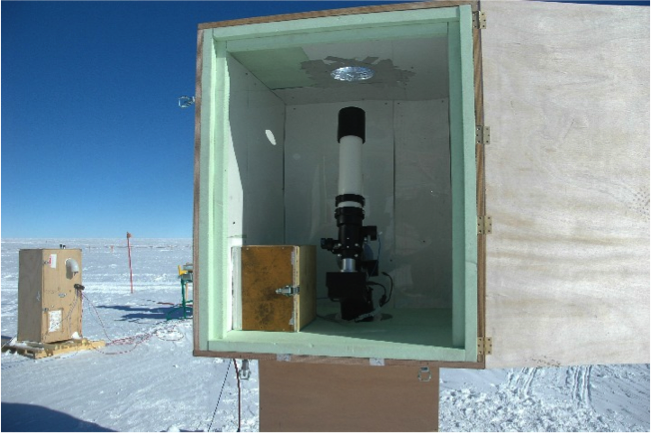}
      \caption{ASTEP South at Dome~C, Antarctica.}
   \label{fig:ASTEPSouthDomeC}
\end{minipage}
\hspace{0.5cm}   
\begin{minipage}[htbp]{7.5cm}
   \centering
   \includegraphics[width=7.5cm]{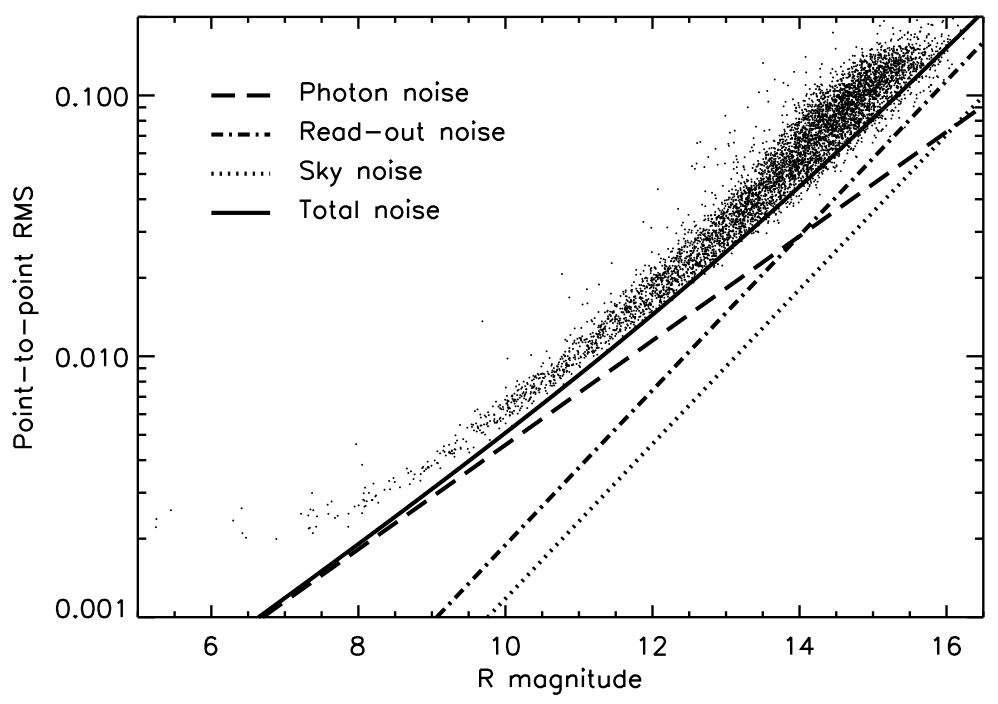}
   \caption{Point-to-point RMS of the June 21, 2008 lightcurves as a function of instrumental magnitude. The RMS is calculated over the 30 s exposure time. The photon noise is represented by a dashed line, the read-out noise by a dash-dot line, the sky noise by a dotted line, and their quadratic sum by a plain line.}
   \label{fig:rms-diagram 2008-06-21}
\end{minipage}
\end{figure}

Figure~\ref{fig:rms-diagram 2008-06-21} shows the point-to-point RMS over 30 s for the 1-day lightcurves of June 21, 2008. Each flux measurement is compared only to its neighbor: the long-term variations are not taken into account. In each lightcurve, we remove outliers departing from the mean by more than 3.5 times the standard deviation. This represents 6\% of the data on average. We obtain a point-to-point RMS of 2.5 mmag at $R=8$, 6 mmag at $R=10$, 1.8\% at $R=12$, and 6\% at $R=14$. This point-to-point RMS is compatible with the expected photon noise, read-out noise, and sky background noise, which are high due to the small size of ASTEP South and the short exposure time. However, these statistical noises should be considerably reduced by binning the huge amount of collected data. \\

\section{First planet candidate: an eclipsing F-M binary}

Lightcurves from the 2008 season are detrended using TFA (\cite{Kovacs2005}) with the method described in \cite{Szulagyi2009}, and periodic signals are searched with BLS (\cite{Kovacs2002}). A first transit candidate is identified on a F dwarf star of coordinates $RA$=18:30:56.777, $DEC$=-88:43:17.01 (J2000) and magnitude $V=10.12$ (instrumental magnitude $R=9.85$). This candidate was also identified by \cite{Wang2011} with the CSTAR instrument at Dome A. We then gather our data from the 2008 and 2009 seasons. We use only the good images, where at least 1/5 of the stars are detected in the field of view (compared to the typical number of stars detected under favorable observing conditions). This represents 73.5\% of the data, or 227633 images. The 2-season lightcurve of this candidate is calibrated with reference stars. A residual variation of 1-sidereal day period appears due to an imperfect flat-fielding. To correct for this, we fold the lightcurve over 1 sidereal day, bin it into 100 points with a median filter, and divide the folded lightcurve by this binned lightcurve. This correction is applied for each season independently. Because residual trends remain, we apply a low-frequency variation correction on the lightcurve of each day by fitting and subtracting a 2\nd order polynomial (the in-transit data points are excluded from the fit). The RMS over 30 s for this 2-season lightcurve is 1.1\%. This is about twice greater than the 1-day point-to-point RMS at that magnitude, indicating that trends still remain in the lightcurves.

\begin{figure}[htbp]
   \centering
   \includegraphics[width=10cm]{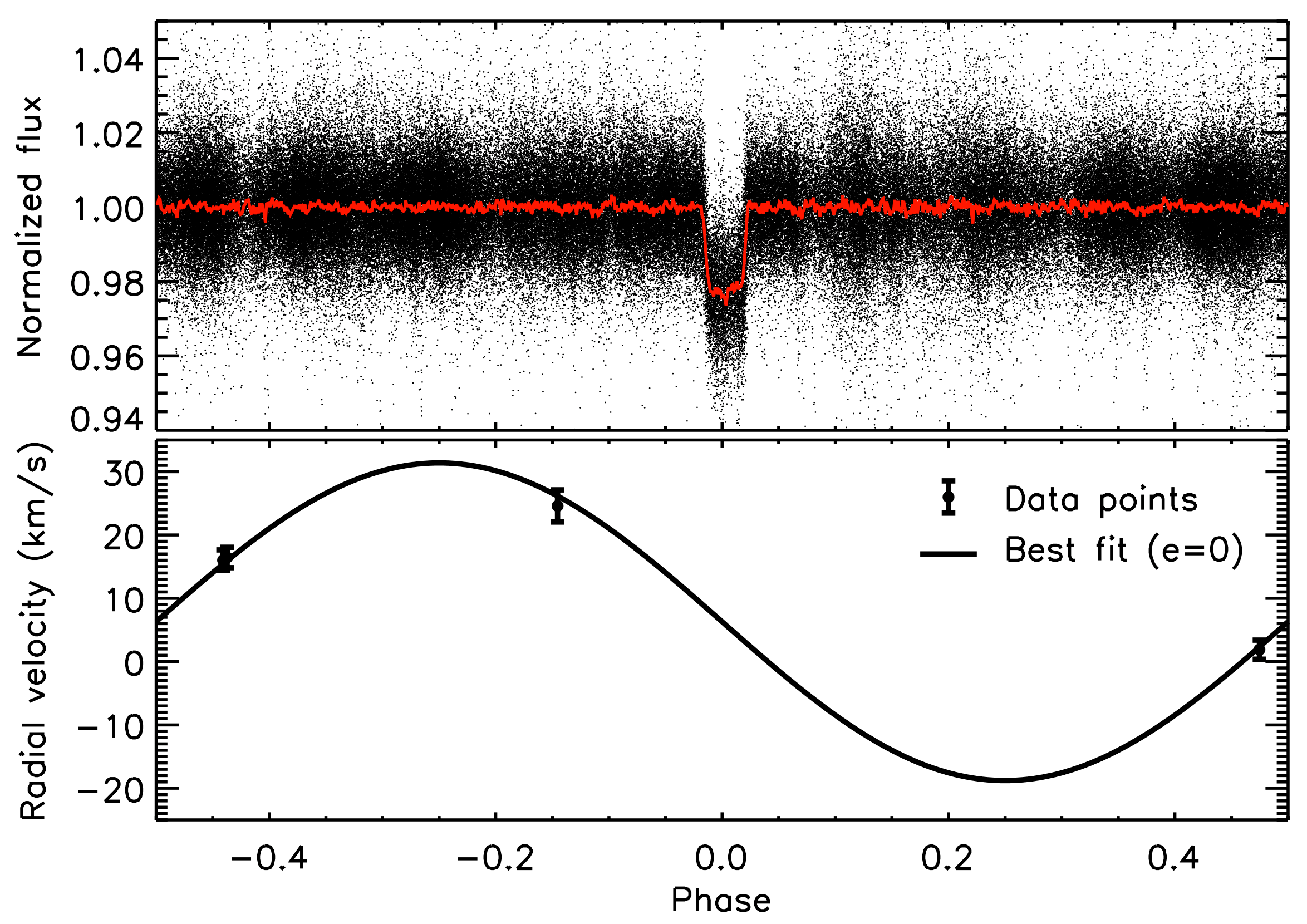}
   \caption{Top: 2-season lightcurve of the star $RA$=18:30:56.777, $DEC$=-88:43:17.01 (J2000), folded at a period $P=9.927$ days (black dots), and binned into 1000 points (red line). The instrumental magnitude is $R=9.85$. The binned lightcurve has RMS of 1.09 mmag. Bottom: Radial velocity measurements (filled circles), and best fit assuming no eccentricity (plain line). This system is an eclipsing F-M binary.}
   \label{fig:candidate}
\end{figure}

Twenty transits are present in the data that we analyzed: 7 in 2008 and 13 in 2009. Most of them are partial transits because of their long duration (\simi10 hours). The period $P$ is calculated using BLS; we find $P = 9.927\pm0.003$ days. The transit time reference is $HJD_0 = 2455060.10 \pm 0.01$. We then fold the lightcurve at the period $P$ and bin it into 1000 points (Figure~\ref{fig:candidate}). The transit depth $\rho$ is calculated using this binned lightcurve; we find $\rho = 2.25\pm0.18$\%. For comparison, \cite{Wang2011} derive a period $P=9.916$ days and a depth $\rho=17$ mmag (1.55\%), from a smaller number of transits and using data from 2008 only. Our binned lightcurve has a RMS of 1.09 mmag, for this 9.85 magnitude star. This precision shows the quality of Dome C for continuous photometric observations. The theoretical photon noise limit is however 0.29 mmag. The use of the 4 seasons of data as well as a better detrending should thus yield a sub-millimagnitude precision for this star, which would be unprecedented for a 10 cm instrument. 

Follow-up radial velocity observations were conducted at the ANU 2.3 m telescope located at Siding Spring Observatory, Australia, in January 2012. Four data points were taken, 2 being almost at the same phase. These measurements show an amplitude too large to be caused by a planet (Figure~\ref{fig:candidate}). This system is therefore an eclipsing binary. No more observations were made. Our spectra confirm the F-dwarf nature of the primary star. Assuming no eccentricity ($e = 0$), we find a radial velocity semi-amplitude $K=25.1\pm2.8$ km/s. Assuming a mass $M_{prim}=1.3 \rm\; M_\odot$ for the primary as a standard mass for F-dwarfs, we derive a mass $M_{sec}=0.35\pm0.05 \rm\; M_\odot$ for the secondary. This system is therefore an eclipsing F-M binary. The detection of this 9.9-day period system is however encouraging for the detection of planets with orbital period longer than those usually detected from the ground.

\section{Conclusion}

ASTEP South has been observing the South pole field almost continuously during winters since 2008. Our analysis of the 2008 and 2009 seasons shows a point-to-point RMS that can be explained by mostly a combination of photon noise and read-out noise. Due to the short exposure time and the small size of ASTEP South, this RMS is high, but should be considerably reduced by the large amount of collected data when searching for periodic signals. A 9.9-day period eclipsing F-M binary is found. After combining data from the 2 first seasons, a precision of 1.09 mmag is obtained for this 9.85 magnitude system. A sub-millimagnitude precision should be reached with a better detrending and the use of the 4 seasons of data. This detection and analysis show the quality of Dome C for continuous photometric observations, even with a small instrument such as ASTEP South, and is encouraging for the search of planets with orbital period longer than those usually detected from the ground. In addition to the search for planetary transits, we will also analyze interesting variable stars present in the field of view.


\newpage

\begin{thebibliography}{}

\bibitem[Aristidi \etal\ 2003]{Aristidi2003}
{Aristidi, E., Agabi, A., Vernin, J., \etal\ }, 2003,
\textit{A\&A}, 406, L19

\bibitem[Aristidi \etal\ 2005]{Aristidi2005a}
{Aristidi, E., Agabi, K., Azouit, M., \etal\ } 2005,
\textit{A\&A}, 430, 739

\bibitem[Aristidi \etal\ 2009]{Aristidi2009}
{Aristidi, E., Fossat, E., Agabi, A., \etal\ } 2009,
\textit{A\&A}, 499, 955

\bibitem[Ashley \etal\ 2005]{Ashley2005a}
{Ashley, M.~C.~B., Lawrence, J.~S., Storey, J.~W.~V., \& Tokovinin, A.} 2005,
\textit{EAS Publications Series}, 14, 19

\bibitem[Crouzet \etal\ 2007]{Crouzet2007}
{Crouzet, N., Guillot, T., Fressin, F., Blazit, A., \& the ASTEP~team} 2007,
\textit{AN}, 328, 805

\bibitem[Crouzet \etal\ 2010]{Crouzet2010}
{Crouzet, N., Guillot, T., Agabi, K., \etal\ } 2010,
\textit{A\&A}, 511:A36

\bibitem[Daban \etal\ (2010)]{Daban2010}
{Daban, J.-B., Gouvret, C., Guillot, T., \etal\ } 2010,
\textit{SPIE Conference Series}, 7733

\bibitem[Fossat \etal\ 2010]{Fossat2010}
{Fossat, E., Aristidi, E., Agabi, A., \etal\ } 2010,
\textit{A\&A}, 517, A69

\bibitem[Fressin \etal\ 2005]{Fressin2005a}
{Fressin, F., Guillot, T., Bouchy, F., \etal\ } 2005,
\textit{EAS Publications Series}, 14, 309

\bibitem[Giordano \etal\ 2012]{Giordano2012}
{Giordano, C., Vernin, J., Chadid, M., \etal\ } 2012,
\textit{PASP}, 124, 494

\bibitem[Kenyon \etal\ 2006]{Kenyon2006a}
{Kenyon, S.~L., Lawrence, J.~S., Ashley, M.~C.~B., \etal\ } 2006,
\textit{PASP}, 118, 924

\bibitem[Kov{\'a}cs \etal\ 2005]{Kovacs2005}
{Kov{\'a}cs, G., Bakos, G., \& Noyes, R.~W. } 2005,
\textit{MNRAS}, 356, 557

\bibitem[Kov{\'a}cs \etal\ 2002]{Kovacs2002}
{Kov{\'a}cs, G., Zucker, S., \& Mazeh, T.} 2002,
\textit{A\&A}, 391, 369

\bibitem[Lawrence \etal\ 2004]{Lawrence2004a}
{Lawrence, J.~S., Ashley, M.~C.~B., Tokovinin, A., \& Travouillon, T.} 2004,
\textit{Nature}, 431, 278

\bibitem[Mosser \& Aristidi 2007]{Mosser2007a}
{Mosser, B., \& Aristidi, E.} 2007,
\textit{PASP}, 119, 127

\bibitem[Pont \& Bouchy 2005]{Pont2005a}
{Pont, F., \& Bouchy, F.} 2005,
\textit{EAS Publications Series}, 14, 155

\bibitem[Rivet \etal\ (2012)]{Rivet2012}
{Rivet, J.-P., \etal\ } 2012,
\textit{IAUS288 Proceedings}, these proceedings

\bibitem[Wang \etal\ (2011)]{Wang2011}
{Wang, L., Macri, L.~M., Krisciunas, K., \etal\ } 2011,
\textit{AJ}, 142, 155

\bibitem[Szul{\'a}gyi \etal\ (2009)]{Szulagyi2009}
{Szul{\'a}gyi, J., Kov{\'a}cs, G., \& Welch, D.~L.} 2009, 
\textit{A\&A}, 500, 917 

\bibitem[Yuan \etal\ 2008]{Yuan2008}
{Yuan, X., Cui, X., Liu, G., \etal\ } 2008,
\textit{Proc. SPIE}, 7012, 70124G

\bibitem[Zhou \etal\ 2010a]{Zhou2010a}
{Zhou, X., Wu, Z., Jiang, Z., \etal\ } 2010a,
\textit{Res. Astron. Astrophys.}, 10, 279

\bibitem[Zhou \etal\ 2010b]{Zhou2010b}
{Zhou, X., Fan, Z., Jiang, Z., \etal\ } 2010b,
\textit{PASP}, 122, 347



\end{thebibliography}
\end{document}